%% file: DSA1D.tex
\documentclass[a4paper,11pt]{article}
\pdfoutput=0 

\usepackage{jcappub} 

\usepackage[T1]{fontenc} 

\usepackage{dcolumn}
\usepackage{bm}
\usepackage{epsf}
\usepackage{verbatim}
\usepackage{amsmath}
\usepackage{amssymb}
\usepackage{color}
\usepackage{verbatim} 
\usepackage[normalem]{ulem} 
\usepackage{mathtools}
\usepackage{amsfonts}

\usepackage{xcolor}
\definecolor{darkgreen}{rgb}{0.0,0.5,0.0}

\usepackage[makeroom]{cancel}

\input{Definitions.tex}

\newcommand{\myz}{{\zeta}}
\newcommand{\mye}{{\epsilon}}

\title{\boldmath Analytic study of 1D diffusive relativistic shock acceleration}

\author{Uri Keshet}

\affiliation{Physics Department, Ben-Gurion University of the Negev, POB 653, Be'er-Sheva 84105, Israel; ukeshet@bgu.ac.il}

\abstract{
Diffusive shock acceleration (DSA) by relativistic shocks is thought to generate the $dN/dE\propto E^{-p}$ spectra of charged particles in various astronomical relativistic flows.
We show that for test particles in one dimension (1D), $p^{-1}=1-\ln\left[\gamma_d(1+\beta_d)\right]/\ln\left[\gamma_u(1+\beta_u)\right]$, where $\beta_u$ ($\beta_d)$ is the upstream (downstream) normalized velocity, and $\gamma$ is the respective Lorentz factor.
This analytically captures the main properties of relativistic DSA in higher dimensions, with no assumptions on the diffusion mechanism.
Unlike 2D and 3D, here the spectrum is sensitive to the equation of state even in the ultra-relativistic limit, and (for a J{\"u}ttner-Synge equation of state) noticeably hardens with increasing $1<\gamma_u<57$,
before logarithmically converging back to $p(\gamma_u\to\infty)=2$.
The 1D spectrum is sensitive to drifts, but only in the downstream, and not in the ultra-relativistic limit.
}

\begin{document}
\maketitle
\flushbottom



\section{Introduction}
\label{sec:Intro}

Diffusive (Fermi) shock acceleration (DSA) is believed to be the mechanism responsible for the production of non-thermal, high-energy distributions of charged particles in collisionless shocks found in diverse astronomical systems (for reviews, see Refs. \cite{Blandford_Eichler_87, MalkovDrury01, TreumannEtAl09Review}).
Particle acceleration is identified in both non-relativistic and relativistic shocks, examples of the latter including shocks in gamma-ray bursts \citep[GRBs; \eg][]{KumarZhang15Review}, pulsar wind nebulae, active galactic nuclei, X-ray binaries \citep[\eg][]{RomeroEtAl17}, and type Ibc supernovae (for a review, see Ref. \cite{PelletierEtAl17}).
For recent reviews of DSA in relativistic shocks, see Refs. \cite{BykovEtAl12Review, SironiEtAl15_review}. 

Collisionless shocks in general, and their particle acceleration in particular, are mediated by electromagnetic waves, and are still not generally understood from first principles. No present analysis self-consistently calculates the long-term generation of waves and the wave-plasma interactions. The particle distribution $f$ is often evolved by adopting some ansatz for the scattering mechanism (\eg isotropic diffusion in velocity angle) and neglecting wave generation and shock modification by the accelerated particles, in the so-called test-particle approximation. This phenomenological approach proved successful in accounting for non-relativistic shock observations. For such shocks, DSA predicts a power-law energy spectrum, $df/dE\propto E^{-p}$, where $p\simeq(r+2)/(r-1)$ is a function of the shock compression ratio $r$ \citep{Krymskii_1977, AxfordEtAl77, Bell_1978, BlandfordOstriker78}. For strong shocks in an ideal gas of adiabatic index $\Gamma=5/3$, this gives $p=2$, in agreement with observations.

Analyses of GRB afterglow observations suggested that the ultra-relativistic shocks involved produce high-energy distributions with $p=2.2\pm0.2$ \citep{Waxman97, BergerEtAl03}.
A similar, $p\simeq 2.2$ spectrum is inferred in other astronomical systems, for example the injection of particles radiating above the UV in the Crab nebula \citep{MarcowithEtAl16}.
Such evidence triggered several studies of test-particle DSA in relativistic shocks, with $p$ calculated for a wide range of $\gamma$, various equations of state, and several scattering mechanisms.

DSA analysis is more complicated when the shock is relativistic, mainly because $f$ becomes anisotropic.
Most studies of the problem have thus used Monte-Carlo simulations and various other numerical techniques \citep[\eg][]{EllisonEtAl90, BednarzOstrowski98, Kirk_2000, AchterbergEtAl01, Vietri03, BlasiVietri05}. For isotropic, small-angle scattering in the ultra-relativistic shock limit, where upstream (downstream) fluid velocities normalized to the speed of light $c$ approach $\beta_u\simeq1$ ($\beta_d \simeq 1/3$), spectral indices $p=2.22\pm0.02$ were indeed found \citep{HeavensDrury88, BednarzOstrowski98, Kirk_2000, AchterbergEtAl01}.
An approximate expression for the spectrum \citep{Keshet_2005},
\begin{equation} \label{eq:p_iso_app}
p = (\beta_u + 2\beta_d - 2\beta_u \beta_d^2 + \beta_d^3) / (\beta_u - \beta_d)\coma
\end{equation}
was shown to agree with numerical results for different equations of state and for a wide range of shock parameters; in particular, it yields $p = 20/9\simeq 2.22$ in the ultra-relativistic limit.

However, the spectrum was found to be sensitive to the angular dependence of the diffusion function, especially in the downstream and at certain angles \citep{Keshet06}, and it is unclear if the isotropic small-angle scattering assumption behind Eq.~(\ref{eq:p_iso_app}) and most numerical studies is justified.
Ab-initio particle in cell (PIC) simulations have successfully demonstrated the onset of particle acceleration \citep{spitkovsky_08b, sironi_spitkovsky_09, sironi_spitkovsky_11a} and the tightly coupled process of magnetic structure growth \citep{keshet_09, SironiEtAl13}.
However, computational limitations prevent such simulations from resolving the three-dimensional (3D) structure of the shock with evolved electromagnetic modes and saturated non-thermal particle distributions, so the steady-state diffusion function remains poorly constrained.
Such studies rely mostly on 2D (spatially; with two or three momentum dimensions), and in some cases even on 1D \citep[\eg][]{dieckmann_06, OkaEtAl11, RajawatSengupta16} simulations, in order to boost the dynamical range; this is useful for studying those effects which are not inherently 3D.

The Fermi acceleration process itself has been studied in 1D previously, but mainly in the Fermi-Ulam model \citep[\eg][]{ZaslavskiiChirikov65,  Dolgopyat08, OliveiraEtAl11}, and in non-relativistic shocks \citep{ZhuEtAl09}, usually with more than one dimension in momentum space.
Here we study DSA for an arbitrary relativistic shock in an effective 1D model; the three-dimensional particle diffusion is thus represented by 1D particle scattering.
This has two main advantages over studies in higher dimensions: (i) the results are analytically tractable; and (ii) there is no need for an unjustified diffusion ansatz.
The results are qualitatively similar to those of higher dimensions, shedding light on the qualitative behavior of the latter, and possibly constraining their ultra-relativistic limit.
In addition, our analysis is directly applicable to one dimensional simulations, and is useful as a rigorous pedagogical exercise and for code verification.

The structure of the paper is as follows.
We begin by introducing the shock jump conditions, which determine the accelerated particle spectrum, and in particular those arising from a 1D J{\"u}ttner-Synge equation of state, in \S\ref{sec:Jump1D}.
The DSA problem in 1D is set up in \S\ref{sec:OneDim}, and solved in the non-relativistic limit in \S\ref{sec:NonRelLimit}.
The spectrum of particles accelerated in an arbitrary relativistic shock is derived in \S\ref{sec:GenericDerivation}, and studied in the ultra-relativistic limit in \S\ref{sec:UltraRel}.
Drifts, known \citep[\eg][]{decker_88} to significantly affect the spectrum, are incorporated in \S\ref{sec:Drifts}.
The results are summarized and discussed in \S\ref{sec:Discussion}.

\section{Relativistic shock jump conditions}
\label{sec:Jump1D}

The shock jump conditions depend on the assumed equation of state, which we write for convenience in the form $P=(\Gamma-1)(\mye-\rho c^2)$.
Here, $P$, $\mye$, and $\rho$ are the proper pressure, energy density, and rest-mass density, respectively, $\Gamma$ is the adiabatic index, and $c$ is the speed of light.
We derive $\Gamma$ and the jump conditions for the standard, J{\"u}ttner-Synge equation of state, below.
But first, we estimate the jump conditions for strong shocks in the non-relativistic and ultra-relativistic limits.

In a non-relativistic gas, $\Gamma\simeq 1+2N_{dof}^{-1}$, where $N_{dof}$ is the number of particle degrees of freedom.
The velocity drop factor at the shock is therefore \citep[\eg][]{LandauLifshitz59_FluidMechanics}
\begin{equation}\label{eq:rNonRel}
r\equiv\frac{\beta_u}{\beta_d} \to \frac{\Gamma+1}{\Gamma-1} \simeq 1+N_{dof} \coma
\end{equation}
where we took the strong shock limit. Here and below, we neglect the back-reaction of the accelerated particles on the adiabatic index, working in the test particle approximation.

In an ultra-relativistic gas, $\Gamma\simeq 1+N_{dof}^{-1}$, so here
\begin{equation}\label{eq:rUltraRel}
r\equiv \frac{\beta_u}{\beta_d} \to \frac{\mye_d+P_u}{\mye_u+P_d} \simeq \frac{\mye_d}{P_d}\simeq \frac{1}{\Gamma-1} \simeq N_{dof} \coma
\end{equation}
again taking the strong shock limit.
For a mono-atomic gas, $N_{dof}=N$, where $N$ is the number of spatial dimensions.
In particular, for an ultra-relativistic shock in a mono-atomic gas in 1D, $r\to N_{dof}=N=1$, indicating no compression or velocity drop across the shock in this limit. Hence, while the asymptotic behavior can be studied, this limit cannot be realized as the shock becomes ill defined.

Equation (\ref{eq:rNonRel}) uniquely determines the spectral index $p$ of test particles accelerated by a non-relativistic strong shock.
Similarly, Eq.~(\ref{eq:rUltraRel}) uniquely determines $p$ for test particles accelerated by a strong shock in the ultra-relativistic limit, but only in 2D and in 3D, and only if the angular diffusion function is specified.
As we show below, this is not the case in a mono-atomic 1D gas.
Here, there is no need to specify the diffusion function, but the precise deviation from the $r\to N=1$ limit strongly influences the spectrum.
We must therefore specify the equation of state and derive the corresponding shock jump conditions.

In the J{\"u}ttner-Synge equation of state \citep{Juttner1911, Synge57}, the adiabatic index $\Gamma$ smoothly interpolates between $1+2N_{dof}^{-1}$ in the non-relativistic limit, and $1+N_{dof}^{-1}$ in the ultra-relativistic limit.
The same arguments made in the 3D case indicate that in 1D, the specific enthalpy is given by $w/\rho c^2=K_2(\myz)/K_1(\myz)$, where $K_n(\myz)$ are the modified Bessel functions of the first kind (McDonald functions) of order $n$, $\myz\equiv mc^2/k_BT$ is the dimensionless inverse temperature, $T$ is the temperature, $k_B$ is the Boltzmann constant, and $m$ is the particle mass.
The adiabatic index is then given by
\begin{eqnarray}
\rmvlrgspc\Gamma  = \frac{w-\rho c^2}{w-\rho c^2-P}
= \frac{1}{1-\myz^{-1}\left[\frac{K_2(\myz)}{K_1(\myz)}-1\right]^{-1}} 
\simeq \begin{cases}
3-\frac{3}{2\myz}+\frac{21}{8\myz^2}+O\left(\myz^{-3}\right) & \mbox{for $\myz\gg 1$ ;} \\
2+\myz+\myz^2\ln \myz+O\left(\myz^{2}\right) & \mbox{for $\myz\ll1$ ,}
\end{cases} 
\end{eqnarray}
in agreement with the non-relativistic and ultra-relativistic limits in Eqs.~(\ref{eq:rNonRel}) and (\ref{eq:rUltraRel}).

The shock jump conditions may be derived in the same method as in the 3D case \citep[\eg][]{KirkDuffy99}.
In a strong shock, the downstream adiabatic index becomes $\Gamma_d \simeq 1+\myz_d^{-1}(\gamma_r-1)^{-1}$.
Here, $\gamma_r\equiv(1-\beta_r^2)^{-1/2}$, where $\beta_r=(\beta_u-\beta_d)/(1-\beta_u\beta_d)$ is the normalized relative velocity of the upstream with respect to the downstream.
The Taub adiabat in the strong shock limit can be written as
\begin{equation} \label{eq:Taub1}
\frac{K_2(\myz_d)}{K_1(\myz_d)} = \frac{\gamma_u}{\gamma_d}= \gamma_r + \frac{1}{\myz_d}
\mbox{ ;}
\end{equation}
\begin{equation} \label{eq:Taub2}
\gamma_u^2 = \frac{w_d^2(\mye_d^2-\rho_d^2c^2)/\rho_d^2c^2}{\mye_d^2-P_d^2-\rho_d^2c^2} = \frac{(\gamma_r^2-1)(1+\myz_d\gamma_r)^2}{(\gamma_r^2-1)\myz_d^2-1} \fin
\end{equation}
For a given shock Lorentz factor $\gamma_u$, Eqs.~(\ref{eq:Taub1}--\ref{eq:Taub2}) can be solved simultaneously for $\gamma_r$ and $\myz_d$. Given this $\gamma_r$, one may now relate $\beta_d$ to $\beta_u$.

\section{DSA set up}
\label{sec:OneDim}

Consider an infinite, planar shock front in 3D, located at shock-frame coordinate $z=0$, with flow in the positive z direction. Relativistic particles with energy $E$ much higher than any characteristic energy in the problem are assumed to diffuse in their direction of motion. This direction is parameterized by the fluid frame $\mu\equiv\cos(\theta)$, where $\theta$ is the polar angle measured between the velocity vector and the positive $z$ axis.

Anticipating the reduction of 3D diffusion to 1D scattering, we use the transport, scattering-convection equation in the form \cite[\eg][]{Vietri03}
\begin{align} \label{eq:ScatConv}
\gamma c  (\beta+\mu) \frac{\partial f}{\partial z}  = \left(\frac{\partial f}{\partial t}\right)_{\mbox{\footnotesize collisions}}
 = \int R({\unit{n}'\to \unit{n}}) f(\unit{n}')\,d\unit{n}' - \int R({\unit{n}\to \unit{n}'}) f(\unit{n})\,d\unit{n}' \coma
\end{align}
where $R({\unit{n}_1\to \unit{n}_2})$ is the fluid-frame rate of collisions that scatter particles from direction $\unit{n}_1$ to $\unit{n}_2$.
We assume, as is customary, that $R$ is separable in the form $R=R_a(z,E)R_b(\unit{n}_1,\unit{n}_2)$.
Equation (\ref{eq:ScatConv}) holds separately in the upstream and in the downstream; we write the upstream/downstream indices $u/d$ only when necessary.
Consider first a constant $R_a$; the general form is addressed in \S\ref{sec:GenericDerivation}.

In 1D, particles propagate only in one out of two directions: either toward the downstream, $\mu=+1$, which we label as the $+$ direction, or toward the upstream, $\mu=-1$, which we label as the $-$ direction.
Diffusion in the direction of motion thus reduces to scattering between these two directions.
It is customary to invoke a symmetry between the scattering probabilities, $R(-\to+)=R(+\to-)$, as an expression of detailed balance \citep[][and references therein]{BlasiVietri05}.
We refer to the case of symmetric scattering as (pure) diffusion, and designate any asymmetry between $R(-\to+)$ and $R(+\to-)$ as a drift.

In the 1D case, the angular distribution thus becomes simple, and the spectrum of the accelerated particles turns out to be independent of the details of their diffusion or scattering.
The absence of a characteristic energy scale implies that a power-law spectrum emerges, as verified numerically \citep{BednarzOstrowski98, AchterbergEtAl01}.
We may therefore write the distribution function as $f=[f_+(z)+f_-(z)]E^{-p}$.

The spectral index $p$ may be derived by solving Eq.~(\ref{eq:ScatConv}) under the following boundary conditions.
Continuity of the accelerated particles across the shock front requires that $f_{u\pm}(z=0)E_u^{-p} = f_{d\pm}(z=0)E_d^{-p}$, where upstream and downstream quantities are related by a Lorentz boost of velocity $c\beta_r$.
The absence of accelerated particles reaching far upstream requires that $f_u(z \to -\infty) = 0$.
The angular diffusion of such particles as they are carried far downstream implies that they eventually isotropize, $f_{d\pm}(E_d,z\to +\infty) = f_\infty
E_d^{-p}$, where $f_\infty>0$ is a constant.

\section{Non-relativistic limit}
\label{sec:NonRelLimit}

There are simpler ways to derive the spectrum of particles accelerated by the shock in the non-relativistic limit, in the present case in 1D.
Consider for example a relativistic particle after $i$ cycles of crossing the shock back and forth, with Lorentz factor $\gamma_i\gg1$ in the upstream frame, $S_u$.
The particle crosses the shock towards the downstream and is then reflected back to the upstream due to elastic collisions in the downstream frame, $S_d$, with some scatterers comoving with the downstream medium.
The balance between the fractional energy gain $g$ of the particle in such a cycle, and its probability $P_{e}$ to escape downstream, fixes the spectrum, as shown in the 3D case by Ref. \cite{Bell_1978}.

After it crosses the shock into the downstream, in $S_d$ the particle has $\gamma_i'=\gamma_r\gamma_i(1+\beta_r\beta_i)\simeq \gamma_i(1+\beta_r)$, where we used $\beta_r\gamma_r\ll 1$ and $\beta_i\gamma_i\gg1$.
Upon returning to the upstream, the particle has an upstream frame $\gamma_{i+1}=\gamma_r^2\gamma_i(1+2\beta_r\beta_i+\beta_r^2)\simeq \gamma_i(1+2\beta_r)$, so the fractional energy gain is $g\equiv(\gamma_{i+1}/\gamma_i)-1\simeq 2\beta_r$.
As in higher dimensions, the downstream distribution $f_d$ is expected to be approximately homogeneous and isotropic in $S_d$.
The flux of particles crossing towards the downstream is then $f_dc/2$, whereas the flux of particles escaping far downstream is $f_d\beta_dc$. The ratio between these two fluxes gives the escape probability, $P_e\simeq 2\beta_d$.

Combining these estimates now yields the spectral index of the accelerated particles,
\begin{equation} \label{eq:P1DNonRel}
p=1-\frac{\ln (1-P_e)}{\ln (1+g)} \simeq 1+\frac{2\beta_d}{2\beta_r}
\simeq\frac{\beta_u}{\beta_u-\beta_d} \fin
\end{equation}
For a non-relativistic, mono-atomic gas in 1D, $\Gamma\simeq 3$, and so $\beta_u\simeq 2\beta_d$, giving the well known, flat, $p\simeq 2$ spectrum also found in higher dimensions.

\section{Arbitrary 1D relativistic shock}
\label{sec:GenericDerivation}

Next, consider an arbitrary relativistic shock.
The two components $f_\pm$ of $f$ are coupled by particle scattering, of fluid-frame rate $R$, according to a 1D version of Eq.~(\ref{eq:ScatConv}),
\begin{align} \label{eq:TwoComp1}
& \gamma c (\beta + 1)\partial_z f_+ = R({-\to+})f_- - R({+\to-})f_+ \, ; \nonumber \\
& \gamma c (\beta - 1)\partial_z f_- = R({+\to-})f_+ - R({-\to+})f_- \fin
\end{align}
In the absence of drifts, which are incorporated in \S\ref{sec:Drifts}, the scattering rate is symmetric, $R\equiv R({-\to+})=R({+\to-})$.
The precise value of this rate, and its spatial and energetic dependencies, are inconsequential for the spectral analysis, as we may absorb them in the definition of a new spatial coordinate, $\tau\equiv \int^z R_a \,dz'/(\gamma c)$.
This yields two simple, coupled equations, which may be compactly written as
\begin{equation} \label{eq:TwoComp2}
 (\beta\pm 1)\pr_\tau f_\pm = f_\mp-f_\pm \fin
\end{equation}

The general solution to these two coupled equations is
\begin{equation}
f_\pm = C_1 +  \frac{C_2}{1\pm\beta} e^{2\beta\gamma^2\tau}\coma
\end{equation}
valid on each side of the shock with different constants $C_1$ and $C_2$, to be determined by the boundary conditions.

In the far downstream, the finiteness of $f_d$ requires that $C_{d2}=0$; the solution throughout the downstream thus becomes homogeneous and isotropic,  $f_{d+}=f_{d-}=C_{d1}$.
In the far upstream, no accelerated particles are assumed to be present, so the $f_{u\pm}(\tau\to-\infty)=0$ boundary condition implies that $C_{u1}=0$.
The particle anisotropy level in the upstream is thus spatially uniform, and given by
\begin{equation}  \label{eq:UpAnisotropy}
f_{u+}/f_{u-}=(1-\beta_u)/(1+\beta_u) \fin
\end{equation}
Hence, the upstream distribution even near the shock is dominated by particles moving toward the upstream, away from the shock.
As one approaches the ultra-relativistic limit, $\beta_u\to1$, the fraction of upstream particles moving toward the downstream becomes exceedingly small. This resembles the behavior of the particle distribution found \citep[\eg Ref.][and Lavi et al., in prep.]{Kirk_2000} in higher dimensions.

\begin{figure*}
\centerline{\epsfxsize=20.5cm \hspace{1.8cm}\epsfbox{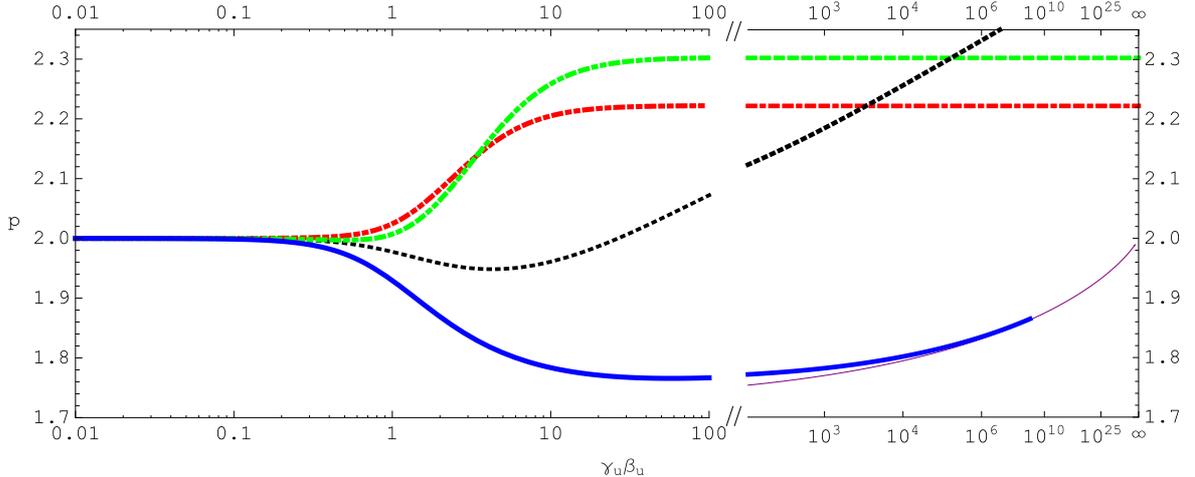}}
\caption{
Energy spectral index $p$ of shock-accelerated particles in 1D (Eq.~(\ref{eq:GenSpec}); thick solid blue curve), 2D (dashed green), and 3D (dot-dashed red), for a strong shock and a J{\"u}ttner-Synge equation of state.
In 2D and in 3D, isotropic diffusion was assumed; the results shown are the analytic approximations of Lavi et al. (in prep.) for 2D, and of Ref. \cite{Keshet_2005} for 3D.
In 1D, the result is sensitive to the equation of state even in the ultra-relativistic limit (where the approximation Eq.~(\ref{eq:BetaDApprox}) for the J{\"u}ttner-Synge equation of state is used; thin purple).
This sensitivity is demonstrated by showing the spectrum also for a different ($\Gamma_d=2+\gamma_u^{-0.4}$) interpolation of the downstream adiabatic index between its non-relativistic and ultra-relativistic limits (dotted black).
}
\label{fig:spec_1D}
\end{figure*}

We may now match the upstream and downstream solutions at the shock, $\tau=0$, taking into account the Lorentz boost between the upstream and downstream frames, such that
\begin{equation}
f_{u\pm}(\tau=0) = \gamma_r^{-p} (1\pm\beta_r)^{-p} f_{d\pm}(\tau=0) \fin
\end{equation}
Matching the two components yields two constraints, $C_{u2}/(1\pm\beta_u)=C_{d1}\gamma_r^{-p}(1\pm\beta_r)^{-p}$.
Solving for $p$ finally yields the spectral index,
\begin{equation} \label{eq:GenSpec}
p
= \frac{\ln\left[\gamma_u(1+\beta_u)\right]}{\ln\left[\gamma_r(1+\beta_r)\right]}
= \frac{1}{1-\frac{\ln\left[\gamma_d(1+\beta_d)\right]}{\ln\left[\gamma_u(1+\beta_u)\right]}} \fin
\end{equation}
This result is shown in Figure \ref{fig:spec_1D}, where it is compared to the spectra derived in 2D and in 3D under the isotropic diffusion ansatz. Notice that Eq.~(\ref{eq:GenSpec}) reduces to Eq.~(\ref{eq:P1DNonRel}) in the non-relativistic limit.

\section{Ultra-relativistic limit}
\label{sec:UltraRel}

In the ultra-relativistic limit in 1D, the compression ratio $r=\beta_u/\beta_d$ asymptotes to unity. Directly introducing this limit into Eq.~(\ref{eq:GenSpec}) leads to a divergence, and so is insufficient for determining the ultra-relativistic limit of the spectrum, $p(\gamma_u\to\infty$).
Therefore, unlike the 2D and 3D cases, in 1D we must specify the equation of state, and thus also the deviation from the limiting value $r(\gamma_u\to\infty)$, in order to derive the asymptotic spectrum.
This is illustrated in Figure \ref{fig:spec_1D} by showing the spectrum for two 1D equations of state, that have the same limiting $\Gamma$ behavior but approach the asymptotic ultra-relativistic limit differently.

Consider the spectrum for the J{\"u}ttner-Synge equation of state, shown in the figure as solid curves.
Remarkably, after the spectrum hardens gradually down to $p(\gamma_u\simeq 57.6)\simeq 1.77$, it turns around and softens back, until converging logarithmically in the ultra-relativistic limit onto the exact same $p\to 2$ value as in the non-relativistic case.
Such spectral hardening followed by softening is found for other equations of state that interpolate between the non-relativistic and ultra-relativistic limits, as illustrated in the figure by the dotted curve, but the limiting spectrum is in general not flat ($p\neq 2$).

Consider the J{\"u}ttner-Synge equation of state in the ultra-relativistic limit. Downstream of the shock, the high temperature implies that $\zeta\to0$.
Solving Eq.~(\ref{eq:Taub1}) to leading order in $\zeta$ yields $\gamma_r = \myz^{-1}-z\ln \myz+O(z)$, diverging in the ultra-relativistic limit in spite of $r\to 1$.
This implies that $\Gamma_d\simeq 2+(\gamma_r-1)^{-1}$, reproducing the anticipated  $\Gamma_d(\gamma_r\to\infty)=2$ of Eq.~(\ref{eq:rUltraRel}).
Expanding Eq.~(\ref{eq:Taub2}) in the ultra-relativistic limit now yields  $\beta_u=1+\myz^4\ln(\myz)/4+O(\myz^4)$ and $\beta_d=1+\myz^2\ln(\myz)+O(\myz^2)$.
These may be combined to give the approximate velocity jump at the ultra-relativistic shock limit,
\begin{eqnarray}\label{eq:BetaDApprox}
\beta_d & \simeq & 1+i(1-\beta_u)^{1/2}W_{-1}[-16(1-\beta_u)]^{1/2} \nonumber \\
& \simeq & 1-[-(1-\beta_u)\ln(1-\beta_u)]^{1/2} \coma
\end{eqnarray}
where $W_k(\xi)$ is the $k$-branch of the Lambert $W$ (product log) function.
The spectrum Eq.~(\ref{eq:GenSpec}) with the approximation Eq.~(\ref{eq:BetaDApprox}) is shown in Figure \ref{fig:spec_1D} as a thin purple curve.
Using the approximation in the second line of Eq.~(\ref{eq:BetaDApprox}), the spectrum in the ultra-relativistic limit becomes approximately
\begin{equation}\label{eq:pApprox}
p(\beta_u\simeq1)\simeq 2\,\frac{\ln[(1-\beta_u)/2]}{\ln\{-(1-\beta_u)/\ln[16(1-\beta_u)]\}} \coma
\end{equation}
which indeed asymptotes to $p(\beta_u\to1)=2$.

\section{Drifts}
\label{sec:Drifts}

Incorporating drifts into the above analysis can be accomplished by allowing for an asymmetric scattering rate, $R\equiv R({-\to+})\neq R({+\to-})$.
Accordingly generalizing Eq.~(\ref{eq:TwoComp2}) now yields
\begin{align} \label{eq:TwoCompDrift}
& \!\!\! (\beta + 1)\partial_\tau f_+ = f_- - q f_+ \, ; \nonumber \\
& \!\!\! (\beta - 1)\partial_\tau f_- = q f_+ - f_- \coma
\end{align}
where the deviation of $q\equiv R({+\to-})/R({-\to+})$ from unity measures the excess drift in the downstream-to-upstream direction.

\begin{figure}[h!]
\centerline{\epsfxsize=8.5cm \epsfbox{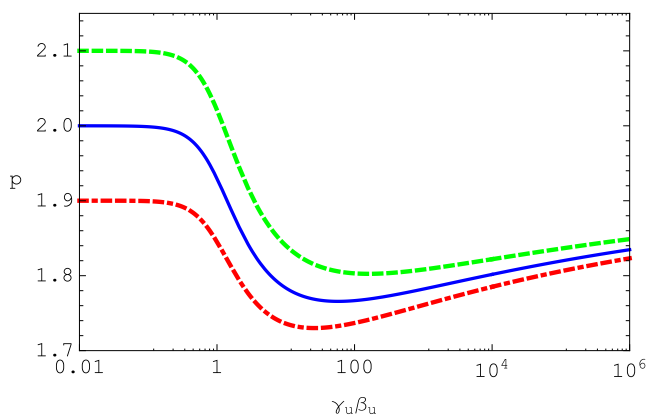}}
\caption{Spectral index of particles accelerated in a strong shock in 1D for a J{\"u}ttner-Synge equation of state, with downstream drift parameter $\xi=0$ (solid blue), $+0.2\beta_d$ (dot-dashed red), and $-0.2\beta_d$ (dashed green).
}
\label{fig:specDrift}
\end{figure}

We may solve these equations analytically if $q$ can be approximated as constant on each side of the shock.
Here, repeating the above considerations leads to the spectrum
\begin{equation} \label{eq:pWithDrift}
p
= \frac{\ln\left[q_d^{-\frac{1}{2}}\gamma_u(1+\beta_u)\right]}{\ln\left[\gamma_r(1+\beta_r)\right]} \\
= \frac{\ln\left[q_d^{-\frac{1}{2}}\gamma_u(1+\beta_u)\right]}{\ln\left[\frac{\gamma_u(1+\beta_u)}{\gamma_d(1+\beta_d)}\right]} \fin
\end{equation}
This result, demonstrated in Figure \ref{fig:specDrift}, is valid only if $q<(1+\beta)/(1-\beta)$ both upstream and downstream, corresponding to a drift which does not strongly push particles toward the upstream.
Otherwise, the upstream boundary condition cannot be satisfied, or the downstream distribution becomes under-determined.
Note that in the ultra-relativistic limit, drifts do not affect the spectrum, \ie $q_d$ can be taken here to unity in Eq.~(\ref{eq:pWithDrift}).

The spectrum in Eq.~(\ref{eq:pWithDrift}) is a function of $q_d$, but is independent of $q_u$ (formally, this is true only as long as $q_u$ is a constant, as assumed above).
The anisotropy $f_{u+}/f_{u-}$ in the upstream is unaffected by either upstream or downstream drifts, and so is still given by Eq.~(\ref{eq:UpAnisotropy}).
The downstream anisotropy, however, depends linearly on the downstream drift, and is now given by $f_{d+}/f_{d-}=1/q_d$.

Equation (\ref{eq:pWithDrift}) reduces in the non-relativistic, weak drift limit to $p\simeq(\beta_u-\xi/2)/(\beta_u-\beta_d)$, where $\xi\equiv q_d-1$ is the downstream drift parameter, here assumed to be small, $|\xi |\ll 1$.
This result can be obtained directly from the non-relativistic derivation in \S\ref{sec:NonRelLimit} by generalizing the escape probability to $P_e\simeq 2(\beta_d-\xi/2)$, as the drift velocity toward the upstream approaches $\xi c/2$ far downstream.

\section{Summary and discussion}
\label{sec:Discussion}

We derived in Eq.~(\ref{eq:GenSpec}) the spectrum of the high-energy tail of relativistic test-particles, accelerated by an arbitrary relativistic shock in 1D. The spectrum, shown in Figure \ref{fig:spec_1D}, is independent of assumptions on the diffusion function, unlike in the 2D and 3D cases.
It reduces to Eq.~(\ref{eq:P1DNonRel}) in the non-relativistic limit, where it yields the familiar flat, $p(\beta_u\to0)=2$ spectrum.
It remains sensitive to the equation of state even in the ultra-relativistic limit, unlike the results in 2D and in 3D, which appear to converge on values independent of the equation of state under the isotropic diffusion ansatz.

The shock jump conditions for the J{\"u}ttner-Synge equation of state, derived for a strong 1D shock in Eqs.~(\ref{eq:Taub1}) and (\ref{eq:Taub2}), yield a spectrum which initially hardens with increasing shock Lorentz factor $\gamma_u$ to $p(\gamma_u\simeq 57.6)\simeq 1.77$, but softens back for faster shocks.
In the ultra-relativistic limit, where the jump conditions and the spectrum are approximated by Eqs.~(\ref{eq:BetaDApprox}) and (\ref{eq:pApprox}), the spectral index logarithmically asymptotes back to the same flat, $p(\gamma_u\to\infty)=2$ value of the non-relativistic case.

Interestingly, some spectral hardening with increasing $\gamma_u$ in the trans-relativistic range was found for a J{\"u}ttner-Synge equation of state also in 3D \citep[][and Arad et al., in prep.]{Kirk_2000}, and possibly also in 2D (Lavi et al., in prep.).
However, in 2D and in 3D this is a weak, barely noticeable effect.

A strong drift toward the upstream would invalidate our analysis, as it can prevent particles from reaching far downstream, or carry them far upstream.
Weak drifts in the upstream have no effect on the 1D spectrum.
The consequences of weak drifts in the downstream are incorporate in Eq.~(\ref{eq:pWithDrift}), and illustrated in Figure \ref{fig:specDrift}.
In general, a downstream drift towards the upstream (downstream) hardens (softens) the spectrum, because it introduces a downstream anisotropy toward the upstream (downstream) and thus lowers (raises) the escape probability.
This effect gradually diminishes with increasing $\gamma_u$, and vanishes altogether in the ultra-relativistic limit.

Test-particle analyses such as presented here do not include non-linear effects caused by the modification of the shock by the accelerated particles themselves, and ignore the injection problem by assuming a pre-existing population of particles that can freely traverse the shock.
Another subtlety is that in such analyses, the velocities $\beta_u$ and $\beta_d$ in the expressions for the spectrum actually pertain to the scattering modes, \ie to the frames in which the particles scatter elastically.
A relative motion between the scattering modes and the fluid, as expected for example in the shock precursor, would accordingly modify the results.

Our analysis provides a rigorous framework to study, at least qualitatively, various effects that are not inherently high dimensional, such as the dependence of the particle distribution upon $\gamma_u$, the role of the equation of state, and the effect of drifts.
For example, the anisotropy of the particle distribution near the shock, given by Eq.~(\ref{eq:UpAnisotropy}), provides a 1D analog of the anisotropic distribution found \citep[\eg][]{Kirk_2000} in 3D.

The emergence of the flat, $p=2$ spectrum, not only in the non-relativistic case, but also in the ultra-relativistic 1D limit, is intriguing.
A flat spectrum, for which the energy in the accelerated particles diverges logarithmically, may serve as an attractor in the ultra-relativistic limit for any dimension.
Such a spectrum was indeed argued in this limit by assuming a microscopically self-similar plasma distribution \citep{Katz_etal_07}.
Thus, is it possible that in higher dimensions, the diffusion function adjusts itself in order to maintain a flat, $p(\gamma_u\to\infty)=2$ spectrum?

\acknowledgments

I thank Y. Nagar, A. Lavi, O. Arad, E. Waxman, A. Heavens, and Y. Lyubarsky for interesting discussions.
This research has received funding from the GIF (grant I-1362-303.7/2016) and the IAEC-UPBC joint research foundation (grant No. 257), and was supported by the Israel Science Foundation (grant No. 1769/15).


\providecommand{\href}[2]{#2}\begingroup\raggedright\endgroup

\end{document}

%% file: Definitions.tex
\newcommand{\ie}{\emph{i.e.} }
\newcommand{\eg}{\emph{e.g.,} }

\newcommand{\be}{\begin{equation}}
\newcommand{\ee}{\end{equation}}
\newcommand{\bea}{\begin{equation*}}
\newcommand{\eea}{\end{equation*}}
\newcommand{\beq}{\begin{equation} }
\newcommand{\eeq}{\end{equation}}
\newcommand{\beqr}{\begin{eqnarray} \nonumber}
\newcommand{\eeqr}{\end{eqnarray}}
\newcommand{\beqrb}{\begin{eqnarray}}
\newcommand{\eeqrb}{\nonumber \end{eqnarray}}
\newcommand{\fin}{\mbox{ .}}
\newcommand{\coma}{\mbox{ ,}}

\newcommand{\unit}[1]{\hat{\mathbf{#1}}}
\newcommand{\pr}{\partial}

\newcommand{\rmvlrgspc}{\!\!\!\!\!\!\!\!\!}

\newcommand{\apjl}{Astrophys. J. Lett.}
\newcommand\apj{{ApJ}}     
\newcommand\mnras{{MNRAS}} 
\newcommand\aapr{Astron Astrophys Rev}
\newcommand\physrep{{Phys.~Rep.}}  
\newcommand{\ssr}{{Space Science Reviews}}
\newcommand{\caa}{{Chinese Astronomy and Astrophysics}}